\documentstyle[12pt]{article}

\begin{document}

\author{T. P. Singh  \\ 
Theoretical Astrophysics Group\\Tata Institute of Fundamental Research\\Homi
Bhabha Road, Mumbai 400005, India}
\title{Some comments on the nature of initial data in spherical collapse}
\maketitle

\begin{abstract}
\noindent Various authors have shown the occurence of naked singularities
and black holes in the spherical gravitational collapse of inhomogeneous
dust. In a recent preprint, Antia has criticised a statement in a paper by
Jhingan, Joshi and Singh on dust collapse. We show that his criticism is
invalid. Antia shows that in Eulerian coordinates a series expansion for the
density of a collapsing Newtonian fluid can have only even powers.
However, he has overlooked the fact that Jhingan et al. have actually used
Lagrangian (comoving) coordinates, and not Eulerian coordinates. As we show,
in Lagrangian coordinates there is no restriction that the density have only
even powers and hence his criticism is invalid. We also point out that an
earlier claim by Antia on the instability of strong naked singularities in
dust collapse is not supported by any concrete analysis, and is hence
incorrect.
\end{abstract}

\noindent A proof or disproof of the cosmic censorship hypothesis is
generally regarded as one of the most important unsolved problems in
classical general relativity. Recent studies of models of gravitational
collapse exhibit examples of formation of both black-holes and naked
singularities, depending on the initial data (for a recent review see, e.g. 
\cite{singh}). Hence the validity of the censorship hypothesis remains an
open problem. In particular, various authors have shown the occurence of
naked singularities in spherical dust collapse \cite{chr}. It turns out that
initial data consisting of only smooth and even functions can result in a
weak naked singularity, whereas if the initial data is allowed to be more
general a strong naked singularity can result \cite{sj}. Jhingan et al. \cite
{jh} discussed the structure of the apparent horizon in dust collapse; this
gives physical insight as to why some initial conditions can lead to a naked
singularity. In this paper the authors also stated that density functions
having odd powers are physically admissible. To clarify this issue, a
comparison was made with the case of the Lame-Emden equation for a polytrope
in Newtonian stellar theory. As is well-known, this equation describes
hydrostatic equilibrium, and it is easily shown from this equation that a
series expansion of the density around the center can have only even powers.
Jhingan et al. pointed out that when the Lane-Emden equation is replaced by
the acceleration equation describing collapse, there is no longer any
restriction that the density function be even. (It should be noted however,
that this comparison is in no way central to the main theme of the paper of
Jhingan et al. which is entirely concerned with dust models, and is no more
than a side remark concerning the possibilities when pressures are included.)

In a recent preprint, Antia \cite{antia} has claimed that this last argument
of Jhingan et al. is incorrect, and that even in the case of a collapsing
fluid, as described by the Newtonian theory, a series expansion for the
density has only even powers. Further, the series expansion for the velocity
can have only odd powers. This is shown by considering a general expansion
in the coupled Euler and continuity equations, written in Eulerian
coordinates. According to Antia, Jhingan et al. did not consider the
continuity equation, and hence arrived at an incorrect conclusion.

The aim of our paper is to point out the error in Antia's reasoning. While
Antia's demonstration holds in Eulerian coordinates, it does not hold in
Lagrangian (i.e. comoving) coordinates, as we show below. Since the
discussion of Jhingan et al. is entirely in comoving coordinates, their
conclusion is hence not affected by Antia's argument, and continues to be
valid. Antia misses the point that the form of functions describing physical
quantities such as density etc. is coordinate dependent and as such changes
with the adoption of coordinates. His statement that we have not included
the continuity equation is completely wrong and misleading. Our analysis
of dust collapse obviously includes the continuity equation. 
Moreover, as we show, our specific argument regarding the 
Lame-Emden equation is unchanged by including the continuity equation, and 
hence in fact this argument does not require the use of the continuity 
equation.

In order to clarify this point, we start by writing the continuity equation
and Euler equation in the Lagrangian coordinates ($t,r$), (see, for
instance, \cite{pod}):

\begin{equation}
\label{cont}\dot \rho +\rho \frac{\dot R^{\prime }}{R^{\prime }}+2\rho \frac{%
\dot R}R=0 
\end{equation}
\begin{equation}
\label{eul}\rho \ddot R=-\rho \frac{Gm(r)}{R^2}-\frac{p^{\prime }}{R^{\prime
}} 
\end{equation}
In these equations, dot stands for the convective time derivative, and prime
stands for a derivative w.r.t. the comoving coordinate $r$. The distance of
a fluid element $r$ from the origin is given by $R(t,r)$. The mass $m(t,r)$
is given by

\begin{equation}
\label{mass}m(t,r)=4\pi \int_0^r\rho R^2R^{\prime }\ dr 
\end{equation}

We assume that the pressure obeys an equation of state $p=p(\rho )$. In
order to check whether Antia's claim holds in Lagrangian coordinates, we
expand the density $\rho (t,r)$ and the distance $R(t,r)$ as a series in the
Lagrangian coordinate $r,$ at an arbitrary time $t$.

\begin{equation}
\label{rho}\rho =\rho _0(t)+\rho _1(t)r+\rho _2(t)r^2+\rho _3(t)r^3+... 
\end{equation}
\begin{equation}
\label{aar}R(t,r)=a_0(t)r+a_1(t)r^2+a_2(t)r^3+... 
\end{equation}
Using these expansions in the Euler equation (\ref{eul}), putting $p^{\prime
}=(dp/d\rho )\rho ^{\prime }$, and assuming the center to be at rest, the
terms at order $r^0$ imply that $\rho _1=0$. (A binomial expansion for $1/R$
and $1/R^{\prime }$ has been used). At order $r$, the continuity equation (%
\ref{cont}) gives $\dot a_1/a_1=\dot a_0/a_0$. Since in general $a_0$ and $%
a_1$ will be non-zero, and are evolving functions of time, we see that there
is no constraint on the series expansion of the velocity $\dot R$. In
general, there will be odd as well as even powers - this is unlike in
Antia's consideration given in Eulerian coordinates. Similarly, since $\ddot
a_0$ and $\ddot a_1$ will in general be non-zero, it does not follow from
the Euler equation that $\rho _3=0$, again unlike what Antia finds. Neither
will the higher odd derivatives vanish. Thus there is no constraint on the
density expansion that it must have only even powers. One could now write
the density as a function of the physical distance. Since there is no
constraint in comoving coordinates, there will be no constraint when the
density is written in terms of the physical distance. The result given by
Antia for Eulerian coordinates is a peculiarity of that particular
coordinate system.

Hence the statement made by Jhingan et al. regarding the series expansion of
the density in comoving coordinates is correct. Antia's criticism is invalid
because he considers expansions in Eulerian coordinates, which are not the
coordinates used by Jhingan et al. It is worthwhile to repeat that the above
discussion for a Newtonian fluid with pressure is in any case rather
academic from the point of view of Jhingan et al., who are actually
considering relativistic dust. On the other hand, Antia tries to create an
impression that our conclusions on dust collapse are incorrect because we
have not used the continuity equation. This is entirely wrong and
misleading. It also demonstrates the irrelevance of the issue raised by him.
Further, it is important to note that for some collapsing systems with
pressure, strong naked singularities actually arise from smooth initial data 
\cite{ori}. Other examples of strong naked singularities are also 
known \cite{peo}. In our opinion, these issues could easily have been 
resolved by dialogue, but the existence of a criticism in the archives 
compels us to reply.

Unfortunately, the preprint by Antia contains a few incorrect statements
which can confuse the reader. Hence we comment on them below.

\noindent (i) Antia writes: ``It may be noted that in all these calculations
the singularity is naked only when it just forms. Irrespective of initial
conditions, gravitational collapse ultimately results in a formation of
event horizon and hence black holes appear to be the only stable end product
of such a collapse''. This statement is incorrect because although the naked
singularity is only a point in the comoving coordinates, it is actually a
null line in the Penrose diagram. Furthermore, an asymptotic observer could
receive a family of null rays emerging from the singularity, for an infinite
period of his time. Besides, since the Cauchy horizon could lie outside the
event horizon in the Penrose diagram, the formation of an event horizon in
such a case could certainly not be guaranteed.

\noindent (ii) Antia criticises an illustration given by Jhingan et al. to
show that the absence of apparent horizon until singularity formation does
not imply the singularity is naked. His criticism is that this illustration
uses a density profile which is non-generic in the space of initial data.
This criticism is completely unfounded. The density profile used by Jhingan
et al. forms a one parameter family of solutions, and even in the possible
case of its being non-generic in a certain sense, it is perfectly reasonable
to use it to demonstrate a conceptual issue.

\noindent (iii) The statements in the last paragraph of Antia's preprint
regarding strength of the singularity do not appear to have any clear
meaning at all. The strength of the naked singularity does not have anything
to do with how the starting epoch is labelled, as is to be expected, since
curvature strength is an invariant quanty.

We would also like to use this opportunity to comment on an earlier paper by
Antia \cite{antia2} where he claims to have shown that strong curvature
naked singularities in dust collapse are unstable. Unfortunately, the paper
does not give any proof to establish the claimed instability. Hence, we
regard the issue of stability of these strong naked singularities as open.
Besides, Antia's paper deals with marginally bound collapse, without
explicitly stating so. Hence it is in major error in claiming the
conclusions to be generally valid for any values of the velocity function;
while that is clearly not what he has shown. Also, the paper incorrectly
states that if the initial density were to increase outwards, the collapse
ends in a black-hole. Actually, in this case the collapse leads to a
shell-crossing (as opposed to shell-focussing) naked singularity, to begin
with, and a discussion of further evolution needs to take into account
possible extendibility beyond the shell-crossing singularity. We will point
out the errors in his paper and discuss these issues elsewhere.

In summary, the objections raised by Antia to our work on dust collapse are
invalid and/or incorrect.

\end{document}